\documentclass[12pt]{iopart}
\usepackage{graphicx}
\usepackage{iopams}
\usepackage{epsf}

\usepackage{color}

\begin{document}

\newcommand{\rem}[1]{{\bf #1}}

















\hfill IPMU-08-0034

\title{Electron and Photon Energy Deposition in Universe}

\author{Toru Kanzaki$^{1}$, Masahiro Kawasaki$^{1,2}$}
\address{$^1$ Institute for Cosmic Ray Research,
 University of Tokyo, Kashiwa 277-8582, JAPAN}
\address{$^2$ Institute for the Physics and Mathematics of the Universe,
 University of Tokyo, Kashiwa 277-8582, JAPAN}

\begin{abstract}

We consider energy deposition of high energy electrons and photons in
universe.  We carry out detailed calculations of fractions of the
initial energy of the injected electron or photon which are used to
heat, ionize and excite background plasma in the early universe for
various ionization states and redshifts.

\end{abstract}



\section{Introduction}
\label{sec:intro}

The energy deposition of fast electrons and high energy photons in
partly ionized plasma is an important issue in the wide range of
physics and astrophysics.  In cosmology, for example, the high energy 
photons and/or
electrons are injected from decay or annihilation of massive particles.
The recent observation of WMAP on Cosmic Microwave Background (CMB) has
shown that dark matter constitutes about 80$\%$ of the total mass in the
universe~\cite{Spergel:2006}. One of the most promising candidates for
dark matter is a supersymmetric particle with mass $O(100)$~GeV. If such
massive particles are annhilate and/or decay, the ejected charged
particles and photons interact with various background particles:
electron, atomic hydrogen and CMB photon, and hence gives the
significant effects on the thermal histrory of the universe. In this
case, to estimate the cosmological effect of dark matter
annihilation/decay  precisely, we need to understand development of
electromagnetic cascade showers induced by primary particles, energy
loss of charged particles and photons, and so on. Besides dark matter,
there are many candidates for inducing electromagnetic showers and
affect the cosmic background plasma in physics beyond the standard model.

Detailed calculations for electrons and photons slowing down in
partially ionized plasma of atomic hydrogen were carried out by
many authors (see references in~\cite{Dalgarno:1999}).  However, these
authors mainly studied the case of injected electron energy up to
keV.  In this paper, we extended these calculations which are valid up
to much higher energy, i.e., $1$~TeV.  The injected high energy
particles produce the cascade showers and lose their energy.  We can
categorize three types of energy loss: {\em heat}, {\em excitation} and
{\em ionization} according to what the energy of the particle is used
for. It is the aim of this paper to derive these quantities
precisely.  They depend on the energy of primary particles as well as
redshift and state of ionization.  There are two reasons why it
is difficult to derive them.  One is the large number of interactions
involved with the energy degradation.  The other is connection between
the energy degradation of charged particles and that of photons.  For
example, high energy electrons can produce photons as much the same
energy through inverse Compton scatterings with CMB photons.  In other
words, we can not calculate the evolution of the energy of charged
particles and photons separately but simultaneously.

In section~\ref{sec:numerical}, we show the numerical method to
calculate the energy degradation of the primary particles.  In
section~\ref{sec:individual}, the evolutions of the energy of primary
electron, photon and positron are presented.  The results of the energy
degradation of the primary particles is shown in
section~\ref{sec:results}.  In section~\ref{sec:conclusion}, we
summarize the results.

\section{Numerical Method}
\label{sec:numerical}

For incident electrons and photons, there are many interactions which
contribute to energy degradation.  It is convenient to divide these
processes into two groups.  One is the group characterized by losing only
a very small portion of energy in one collision (continuous
loss).  The other is the group characterized by being likely to lose a
significant portion of  energy in one collision ('catastrophic'
loss~\cite{Ginzburg:1964}).  In the case of the latter, it is necessary
to use an integro-differential equation to calculate electron energy
spectrum.

Let $E_1, E_2, \cdots ,E_N$ be a discrete set of energies of particles
($E_i < E_j$ for $i< j$) and $N_p(E_s)\Delta E_s$ be the number of
particles with energy between $E_s-\Delta E_s/2$ and  
$E_s+\Delta E_s/2$.  The accuracy of
the numerical method is limited by the bin size ($\Delta E_i$).  Since
we consider very large energy region ranges from $10$ eV to $100$ GeV,
the bin sizes are taken to increase as energy so that $\Delta E_i/E_i$
become constant.  The particle energy spectrum is given by
\begin{eqnarray}
  \frac{\partial N_p(E_s)}{\partial t} 
   & = & -\frac{1}{\Delta E_s}\left[\frac{-dE}{dt}\right]N_p(E_s)
  -N_p(E_s)\sum_{i<s}P(E_s,E_i) \nonumber \\
   & + &\sum_{i>s}N_p(E_i)P(E_i,E_s)+Q(E_s),
    \label{eq:energy_spectrum}			
\end{eqnarray}
where $P(E_i,E_j)dt$ is the probability that a particle with energy
$E_i$ undergo a collision causing it to lose energy and have energy
$E_j$ in time $dt$, and $Q(E_i)$ represents sources and sinks of
particles corresponding to possible production, annihilation or gradual
leakage from the energy range that we consider~\cite{Blumenthal:1970}.
The first term of the R.H.S. of Eq.~(\ref{eq:energy_spectrum})
represents continuous loss and the second and third terms represent
outflow and inflow caused by catastrophic loss.

In this paper, we are mainly interested in how much initial particle
energy convert to heat, excitation and ionization.  For convenience, we
define $\chi_h(E)$, $\chi_e(E)$ and $\chi_i(E)$ as fractions of the
initial energy $E$ which go to heat, excitation and ionization,
repectively.\footnote{
We frequently use ``fraction of heat'',``fraction of excitation'' and
``fraction of ionization'' to refer $\chi_h$, $\chi_e$ and $\chi_i$,
respectively. Please do not confuse $\chi_i$ with ``ionization
fraction'' $x_e$ which is the fraction of ionized hydrogens.}

Suppose that we know already $\chi_h(E_i)$, $\chi_e(E_i)$ and
$\chi_i(E_i)$ with energy less than $E_s$ and consider a particle with
initial energy $E_s$.  Since there are no particle with energy more than
$E_s$, the energy degradation is characterized by the first term
of the R.H.S. of Eq.~(\ref{eq:energy_spectrum}) (continuous loss), the
second term (catastrophic loss) and the last term (annihilation).  
Here we only consider that annihilation of the primary particle 
contributes to $Q(E_s)$. 
All these terms are in proportion to $N_p(E_s)$, so we define 
the following number loss function
\begin{eqnarray}
  L(E_s) = \frac{1}{\Delta E_s}\left[\frac{-dE}{dt}\right]
  +\sum_{i} n_{t}v_{p}\sigma_i(E_s)
\end{eqnarray}
where $n_{t}$ is the number density of target particle, $v_{p}$ is the
particle velocity and $\sigma_i(E_s)$ is the catastrophic and
annihilation cross sections.  From number loss function, we can obtain
the probability that a particle undergoes a particular
collision ``$m$''~\cite{Dalgarno:1971}.  The collision frequency for a
particular continuous loss may be defined by
\begin{eqnarray}
  \nu_m (E_s,E_{s-1}) = \frac{1}{\Delta E_s}\left[\frac{-dE}{dt}\right]_m
\end{eqnarray}
and the collision frequency for a particular catastrophic loss 
``$\alpha$'' and annihilation ``$\beta$'' are given by
\begin{eqnarray}
  \nu_\alpha (E_s,E_i) & = &  n_{t}v_{p}\sigma_\alpha(E_s,E_i),\\
  \nu_\beta (E_s) & = &  n_{t}v_{p}\sigma_\beta(E_s).
\end{eqnarray}
The total collision frequency for the catastrophic loss is given by
\begin{eqnarray}
  \nu_\alpha (E_s)=  \int dE n_{t}v_{p}\sigma_\alpha(E_s,E)
\end{eqnarray}
Then the probability $P(E_s,E_i)$ is written as
\begin{eqnarray}
  P(E_s,E_{i}) & = & 
  \frac{\sum_\alpha \nu_\alpha(E_s,E_i)}{\sum_m \nu_m(E_s,E_{s-1})
  +\sum_\alpha (\nu_\alpha(E_s)+\nu_\beta(E_s))} 
  ~~ (i \neq s-1 ), \\[0.7em]
  P(E_s,E_{s-1}) & = & 
  \frac{\sum_m \nu_m(E_s,E_{s-1})+\sum_\alpha \nu_\alpha(E_s,E_{s-1})}
  {\sum_m \nu_m(E_s,E_{s-1})+\sum_\alpha (\nu_\alpha(E_s)+\nu_\beta(E_s))} .
\end{eqnarray}
Combining these probabilities with the data about $\chi_h(E)$,
$\chi_e(E)$ and $\chi_i(E)$ with energy less than $E_s$, $\chi_h(E_s)$,
$\chi_e(E_s)$ and $\chi_i(E_s)$ can be obtained.  Please notice that the
definition of the frequency for continuous loss depends on the size of
the energy bin but final result is independent of it. The reason is
as follows. When the all bins are divided into halves, the probability
$P(E_s,E_i)$ decreases by fifty percent if the bin size is small
enough.  This is because the collision frequency for continuous loss is
much larger than that for catastrophic loss in this case.  However, the
effect of the decrease of the probability is cancelled by the increase
of the bin numbers.  We have checked that our results are independent
of the size of the energy bin.

\section{Individual Evolution}
\label{sec:individual}

\subsection{Electron}

\begin{figure}[t]
 \begin{center}
  \includegraphics[width=0.6\linewidth]{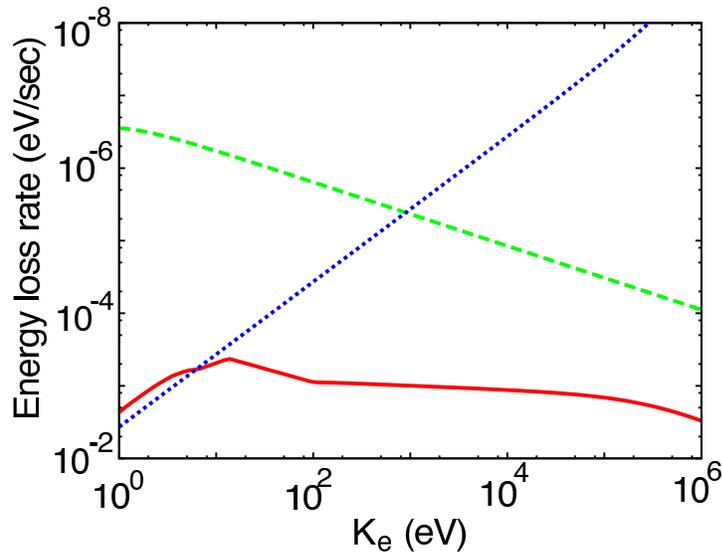}
  \caption{Continuous energy loss rates for electrons with 
  $x_e = 10^{-2}$ and $1+z=1000$. 
  Solid line represents elastic collision with atomic hydrogen,
  dashed line represents Coulomb loss with background electron 
  and dotted line represents inverse Compton scattering with CMB photon.}
  \label{fig:energy_loss_electron}
 \end{center}
\end{figure}

\begin{figure}[t]
 \begin{center}
  \includegraphics[width=0.6\linewidth]{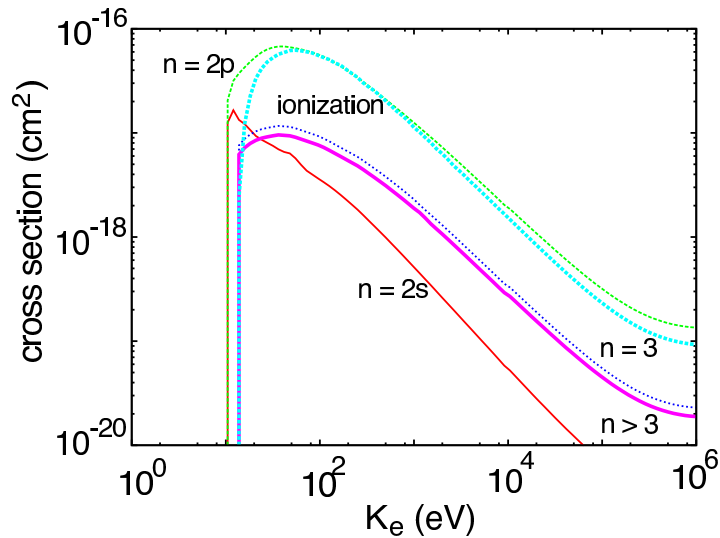}
  \caption{Cross sections for electron impact excitation 
  and ionization of H.}
  \label{fig:cross_section_electron}
 \end{center}
\end{figure}

For incident high energy electrons, 
there are many interactions which contribute to
energy degradation: elastic collision with atomic hydrogen, excitation
and ionization of atomic hydrogen, Coulomb loss with background
electrons and inverse Compton scattering with background photons (CMB
photons).  The cross sections and energy loss rates of these
interactions are described  in~\ref{app:rel} and shown
Figs.~\ref{fig:energy_loss_electron}
and~\ref{fig:cross_section_electron}.  Hereafter, we use electron
kinetic energy ($K_e$) instead of electron energy ($E_e=K_e+m_e$) for
convenience.  In this paper, we have used 4 level (2s, 2p, $n=3$ and
$n>3$) approximation in considering electron impact excitation.  From
Fig.~\ref{fig:energy_loss_electron} it is found that energy loss is
dominated by collisions with background electron (Coulomb loss) at
low energy and inverse Compton scattering off CMB photon at high energy.
The reason is simple.  At low energy, the average energy loss of
electron is roughly $\Delta E_e \propto \beta^2E_{CMB}$ for one inverse
Compton scattering.  So electrons lose very few fraction of their energy
and hence energy loss is dominated by collisions with atomic hydrogen
and background electron.  On the other hand, at high energy, since the
number density of CMB photon is much larger than that of background
electron, energy loss is dominated by inverse Compton scattering.
Energy loss by Coulomb collisions is so efficient that free electrons
with ionization fraction, $x_e$, exceeding $10^{-4}$ have a substantial
influence on the energy degradation~\cite{Dalgarno:1999}.  Thus, as the
ionization fraction increases, the fraction of the initial electron
energy which is converted to heat ($\chi_h(E)$) increases as shown later.

\begin{figure}[t]
 \begin{center}
  \includegraphics[width=0.6\linewidth]{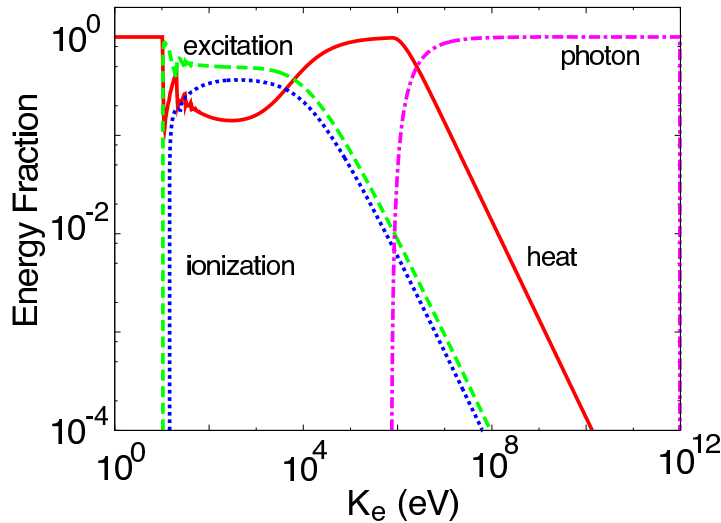}
  \caption{Electron energy degradation with $x_e = 0$ and $1+z=1000$. 
  Solid line represents $\chi_h$, dashed line represents $\chi_{ex}$ 
  and dotted line represents $\chi_i$.
  Dot dashed line represents the fraction of photon energy which energy 
  is larger than Ry.}
  \label{fig:electron_1}
 \end{center}
\end{figure}

\begin{figure}[t]
 \begin{center}
  \includegraphics[width=0.6\linewidth]{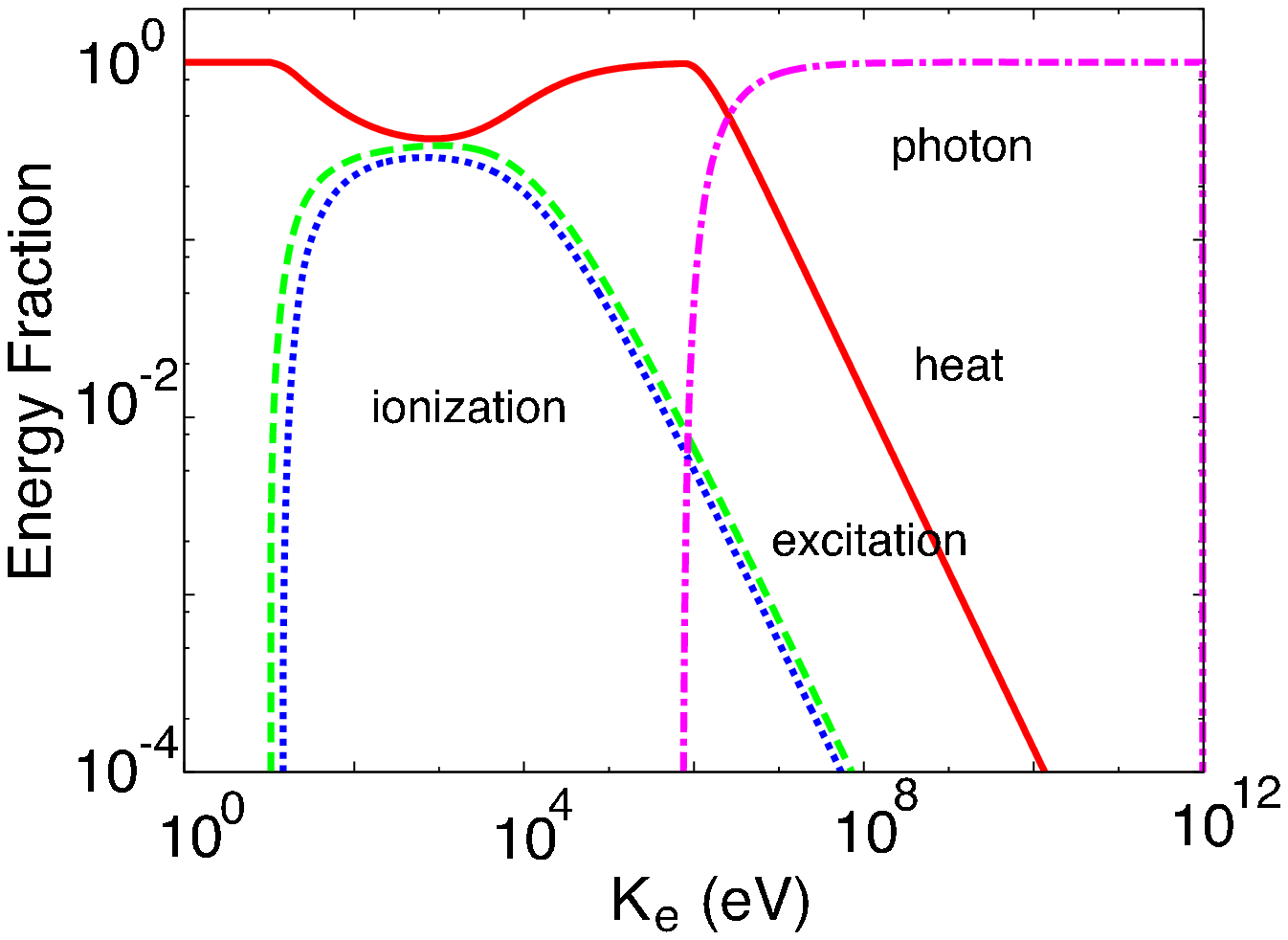}
  \caption{Electron energy degradation with $x_e = 10^{-2}$ and $1+z=1000$. 
  Solid line represents $\chi_h$, dashed line represents $\chi_{ex}$ 
  and dotted line represents $\chi_i$.
  Dot dashed line represents the fraction of photon energy which energy 
  is larger than Ry.}
  \label{fig:electron_2}
 \end{center}
\end{figure}

\begin{figure}[t]
 \begin{center}
  \includegraphics[width=0.6\linewidth]{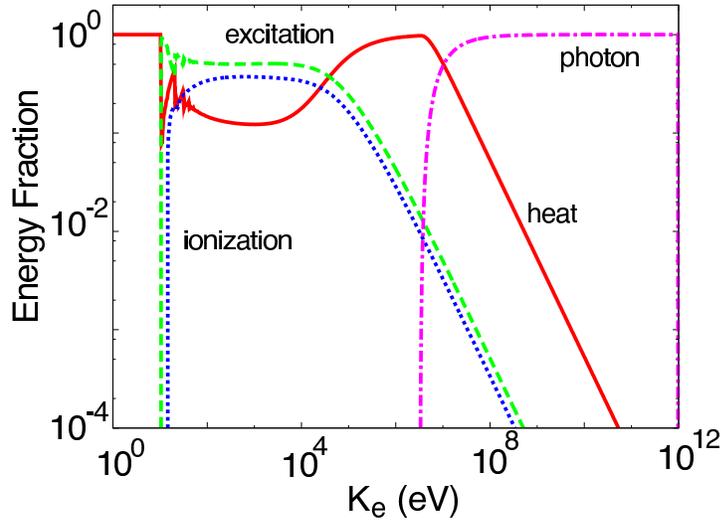}
  \caption{Electron energy degradation with $x_e = 0$ and $1+z=100$. 
  Solid line represents $\chi_h$, dashed line represents $\chi_{ex}$ 
  and dotted line represents $\chi_i$.
  Dot dashed line represents the fraction of photon energy which energy 
  is larger than Ry.}
  \label{fig:electron_3}
 \end{center}
\end{figure}

We calculate the energy degradation of electron following the method
described in the previous section.  There are two free parameters:
ionization fraction $x_e$ and redshift $1+z$.  Ionization fraction is
relevant for collisions with atomic hydrogens and background electrons,
and redshift is mostly relevant for inverse Compton scattering.  For
simplicity, we regard the distribution of CMB photons as monoenergetic
($E_\gamma = 6.34\times10^{-13}(1+z)$ GeV), not blackbody.  Free and
bound proton number densities can be parameterized by baryon-to-photon
ratio $\eta$: $n_p = \eta n_{CMB}$.  We adopt $\eta = 6.1\times
10^{-10}$ from the result of WMAP 3-year
observation~\cite{Spergel:2006}.  We take into account properly
secondary electrons which is produced by electron impact ionization.

Energy degradation for several values of $x_e$ and $1+z$ is shown in
Figs.~\ref{fig:electron_1}-\ref{fig:electron_3}.  Here, we plot
$\chi_h(E)$, $\chi_{ex}(E)$ and $\chi_i(E)$.  In addition, we also plot
the fraction of the initial energy that goes to photons with energy
larger than Ry($=13.6$~eV).  This energy is the threshold energy for
photoionization, and it is granted that these photons also contribute to
heat, excitation and ionization and we will treat them in
Section~\ref{sec:results}.  On the other hnad, photons with energy less
than Ry only heat background electrons with Compton scattering. An
oscillating behavior below $50$~eV in Figs.~\ref{fig:electron_1} and
\ref{fig:electron_3} reflects the nature of discrete energy loss.  This
behavior is not seen in Fig.~\ref{fig:electron_2} since the effect of
Coulomb loss dominates over ionization and excitation.  The common
features in Figs.~\ref{fig:electron_1}-\ref{fig:electron_3} are as
follows: (1) the fractions of heat, excitation and ionization are the
same order for $E_e<10^4$~eV, (2) heat dominates at $10^4<E_e<10^6$~eV,
and (3) finally the energy of photons dominates for $E_e>10^7$~eV.  At
relatively low energy, inverse Compton scattering is inefficient.
Moreover, the cross sections and energy loss rates for momentum loss,
exitation and ionization are almost the same order if 
ionization fraction is
not large, i.e. Coulomb loss is not dominant.  As a consequence,
$\chi_h$, $\chi_{ex}$ and $\chi_i$ are nearly the same order and our
calculation corresponds to the result of~\cite{Dalgarno:1999} up to
$E_e\sim{\rm keV}$.  As electron energy increases, inverse Compton
scattering becomes significant.  If the energy of scattered-up photon is
less than the threshold energy of photoionization, the energy loss due
to inverse Compton scattering converts to heat and the fraction of heat
approaches to unity.  When electron energy further increases and most of
scattered-up photons have enough energy to ionize atomic hydrogen, the
energy of incident electron exclusively converts to the photon energy.

Next we will show how these fractions are dependent on parameters.  To
see the effect of ionization fraction, let us compare
Fig.~\ref{fig:electron_1} with Fig.~\ref{fig:electron_2}.  It can be
seen that the fraction of heat increases at low energy as ionization
fraction increases.  This is because Coulomb collision converts initial
electron energy into heat exclusively. 
Besides,  the collision frequencies
for ionization and excitation are proportional to the number density of
hydrogen atom, $(1-x_e)$, and hence they are alomost independent of 
$x_e$ if $x_e\ll 1$. Thus, $\chi_{ex}(E)$ and $\chi_{i}(E)$ is smaller
as ionization fraction increases.  At high energy, the results are
irrelevant for ionization fraction since inverse Compton scattering
dominates over all other processes.  Comparing Fig.~\ref{fig:electron_1}
with Fig.~\ref{fig:electron_3}, it can be seen that the rise of heat and
photon energy shifts to high energy side as redshift decreases.  This is
because the effect of inverse Compton increases in proportion to the
energy of CMB photon, and hence in inverse proportion to redshift.

\subsection{Photon}

\begin{figure}[t]
 \begin{center}
  \includegraphics[width=0.6\linewidth]{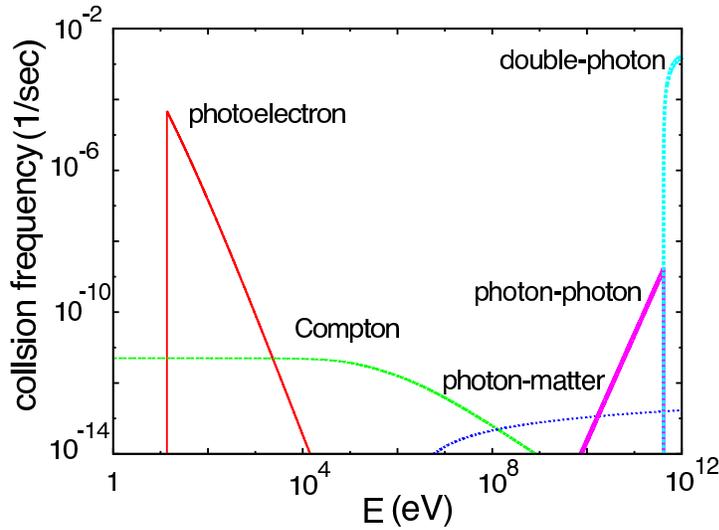}
  \caption{Photon collision frequencies with $x_e = 0$ and $1+z=1000$.}
  \label{fig:frequency_photon}
 \end{center}
\end{figure}


As well as electron, there are many interactions which contribute to
energy degradation for incident high energy photons: photoionization,
Compton scattering with background electrons, pair production in matter,
photon-photon scattering and double photon pair creation.  Here, we
neglect photo-excitation since the resultant excited state immediately
emits photon and goes down to the ground state.  The cross sections and
energy loss rates of these interactions are found in~\ref{app:rel} and
shown in Fig.~\ref{fig:frequency_photon}.

\begin{figure}[t]
 \begin{center}
  \includegraphics[width=0.6\linewidth]{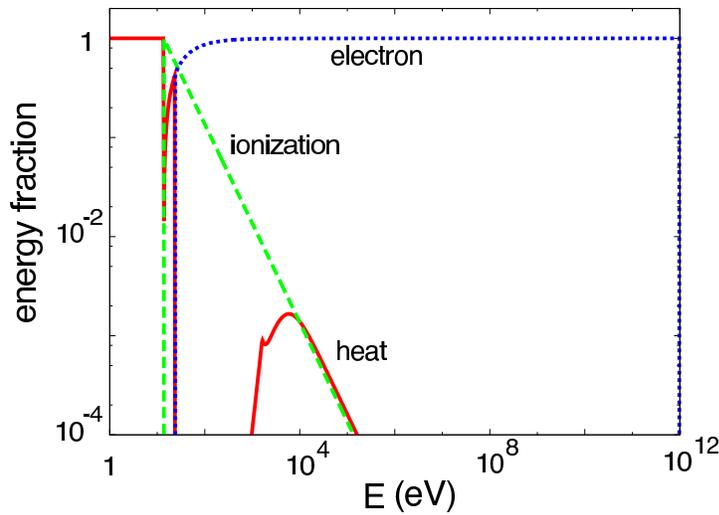}
  \caption{Photon energy degradation with $x_e = 0$ and $1+z=1000$. 
  Solid line represents $\chi_h$, dashed line represents $\chi_{i}$ and
  dotted line represents the fraction of electron energy which energy 
  is larger than 0.75Ry.}
  \label{fig:photon_1}
 \end{center}
\end{figure}

\begin{figure}[t]
 \begin{center}
  \includegraphics[width=0.6\linewidth]{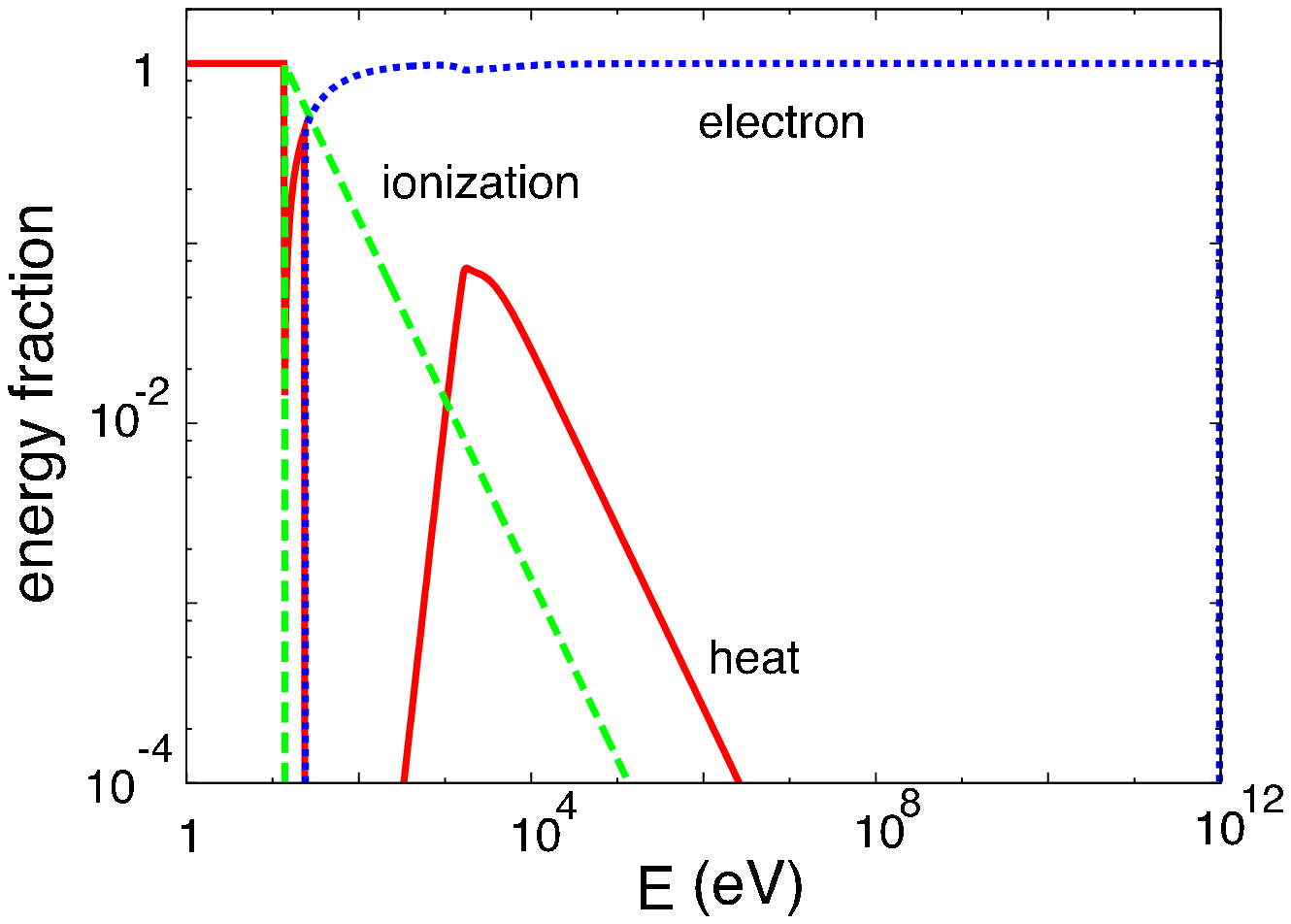}
  \caption{Photon energy degradation with $1-x_e = 10^{-2}$ and $1+z=1000$. 
  Solid line represents $\chi_h$, dashed line represents $\chi_{i}$ and
  dotted line represents the fraction of electron energy which energy 
  is larger than 0.75Ry.}
  \label{fig:photon_2}
 \end{center}
\end{figure}

Energy degradation is plotted for $x_e=0$ and $x_e=0.98$ in
Fig.~\ref{fig:photon_1} and Fig.~\ref{fig:photon_2}.  Here, we plot
$\chi_h$ and $\chi_i$.  In addition, we also plot the fraction of
initial energy which goes to electrons with energy larger than 0.75Ry.
This energy is the threshold for electron impact excitation of atomic
hydrogen. Electrons with lower energy only heat background electrons
through elastic collision with atomic hydrogen and Coulomb loss.  The
sudden falls of $\chi_h$ reflects the nature of discrete energy loss.
The energy for the first fall ($E_\gamma = {\rm Ry}$) corresponds to the
threshold energy of photoionization and the second one ($E_\gamma =
7/4{\rm Ry}$) corresponds to the sum of the threshold energy of
photoionization and that of electron impact excitation. In that case the
energy of the incident photon goes into ionization and the electron
produced in the ionization, and is not used for heating. (Of course the
electron further interacts with background plasma and heats it. This
will be taken account in the next section.) The rise of $\chi_h$ around
$E_\gamma\sim 10^4{\rm eV}$ is  due to Compton scattering.  Unlike the
case of incident electron, the fraction of electron energy dominates
even at low energy.  This is because energetic electrons are produced
through photoionization at low energy and Compton scattering at high
energy.

The increase of ionization fraction causes the increases $\chi_h$
especially at low energy.  This is because photoionization and Compton
scattering determine the degradation of electron energy at low energy.
The effect of photoionization decreases as ionization fraction
increases, and hence $\chi_h$ increases.  
Since the photoionization rate is proportional to $(1-x_e)$,
the effect of photoionization becomes small as $x_e$ approaches $1$, 
which leads to relative enhance of Compton scattering and 
increases $\chi_h$.
Note that we use the the
baryon density instead of the electron density when we calculate the
energy losses due to Compton scattering.  This is because Compton
scattering only becomes important when $E_\gamma$ is sufficiently larger
than Ry and the interaction is insensitive to whether an electron is
bound or not~\cite{Chen:2003}.  As redshift increases, photon-photon
scattering and double photon pair creation become important.  These
processes, however, become dominant at very high energy as seen in
Fig.~\ref{fig:frequency_photon}.  This effect is not appeared until the
energy deposition of electron and positron taken into account.

\subsection{Positron}

Positrons are produced by pair production in electric field of nuclei
and double photon pair creation.  Therefore, we should take the energy
degradation of positrons into account.  Although the energy degradation 
of positrons is almost the same as that of electron, there are
two differences between electron and positron.  One is the sign of its
charge.  The other is the indistinguishability between a primary
electron and a secondary electron in the processes where target is an
electron or an atomic hydrogen.  These differences become less important at
high energy.  For simplicity, we assume positrons lose their energy just
like electrons in this paper.  Besides, positrons finally annihilate
with background electrons through either free annihilation or the
formation and decay of positronium~\cite{Crannell:1976}.  
It depends on the temperature,
density and state of ionization of the background electron which process
is dominant.    Roughly speaking, most of
the positrons undergo annihilation after the significant loss of its
energy.  Therefore, we  assume that a positrons forms a
positronium with a background electron after losing almost all of its
energy and decay.  The positronium annihilates to two photons (each
0.511 MeV) 25\% of the time, and to three photons (each less than 0.511
MeV) 75\% of the time.  The energy spectrum from the three-photon
annihilation is described in~\ref{app:rel}.

\section{Results}
\label{sec:results}


\begin{figure}[t]
 \begin{center}
  \includegraphics[width=0.6\linewidth]{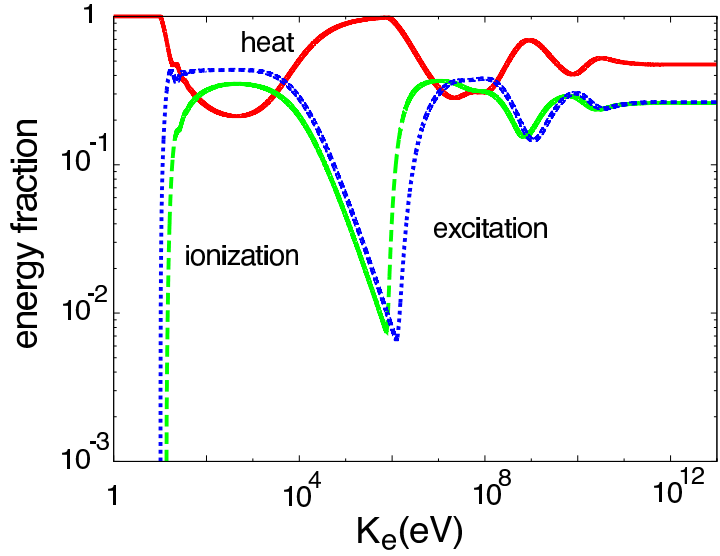}
  \caption{Electron energy degradation with $x_e = 10^{-3}$ and $1+z=1000$. 
  Solid line represents $\chi_h$, dashed line represents $\chi_{i}$ and
  dotted line represents $\chi_{ex}$.}
  \label{fig:electron}
 \end{center}
\end{figure}

\begin{figure}[t]
 \begin{center}
  \includegraphics[width=0.6\linewidth]{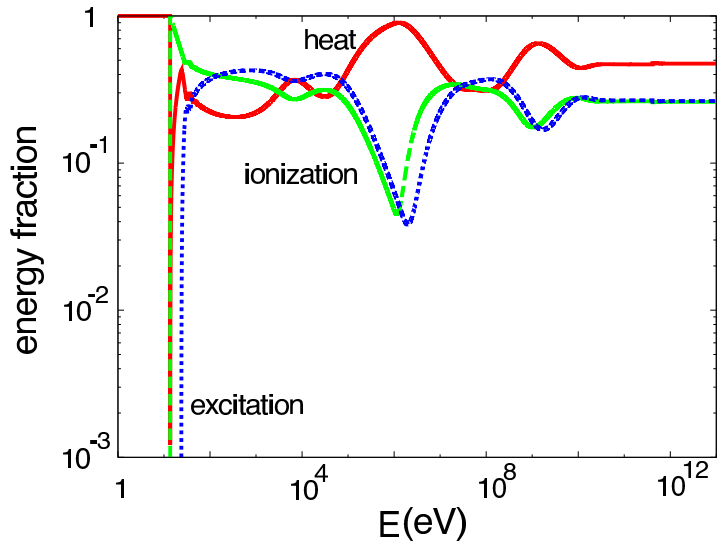}
  \caption{Photon energy degradation with $x_e = 10^{-3}$ and $1+z=1000$. 
  Solid line represents $\chi_h$, dashed line represents $\chi_{i}$ and
  dotted line represents $\chi_{ex}$.}
  \label{fig:photon}
 \end{center}
\end{figure}

As previously mentioned, we should calculate the evolution of the energy
of charged particles and photons simultaneously.  In
Fig.~\ref{fig:electron} and Fig.~\ref{fig:photon}, we show the fraction
of heat, excitation and ionization when the primary particle is electron 
and photon respectively.  Let us examine these figures.  

First, we consider the electron case.  When the energy of the electron
is small ($K_e < 10^4$ eV) and the ionization frcraction is not large, 
the fractions of heat, excitation and ionization are roughly
the same order.  This is because relative smallness of ionization
fraction makes the effect of Coulomb loss comparable with that of
excitation and ionization of atomic hydrogen.  The fraction of heat
increases as ionization fraction increases and vice versa.  For larger
electron energy ($K_e >10^4$ eV), the dominant energy loss mechanism is
the inverse Compton scattering.  If the electron is non-relativistic, the
energy of scattered-up photons is so small that these photons only
contribute to heating of the background particles.  As a consequence,
the fraction of heat reaches near unity.  When an electron is
relativistic, the energy of scattered-up photons exceeds the threshold
energy of ionization of atomic hydrogen.  In this case, it is necessary
to estimate the contribution of these photons.  It is seen that the
fractions of excitation and ionization are a little larger than that of
heat with low photon energy in Fig.~\ref{fig:photon}.  For this reason,
all of these fractions becomes almost the same amount.  The oscillating
structure of the $\chi$'s around $K_e \simeq10^8 \sim 10^{10}$ eV
reflects that of scattered-up photon around $E_\gamma \simeq10^5 \sim
10^{7}$~eV.  When the energy of electron becomes ultra-relativistic,
these oscillating structure is averaged out and vanishes.

Next, let us consider the photon case.  When the energy of the incident
photon is small ($E_\gamma < 10^3$ eV), the fractions of excitation and
ionization are larger than that of heat.  This is because the dominant
energy loss mechanism is photoelectron effect unless ionization fraction
is very close to unity.  In this case, a photon ionizes an atomic
hydrogen and emits a photoelectron whose energy is almost the same as
the primary photon.  Therefore, the behavior of photons is very similar
to that of electrons at low energy.  As photon energy increases, Compton
scattering becomes the dominant energy loss.  When the photon energy is
larger than the electron mass, the forward scattering becomes dominant
and the energy of the recoil electron approaches to that of the photon.
In other words, Compton scattering produces a recoil electron with
energy same as the incident photon.  Therefore, the behavior of photons
traces that of electrons in this energy region.

When photon energy is larger than $10^8$ eV, the effect of
pair-production in matter, photon-photon scattering and double photon
pair creation can not be neglected.  However, it is hard to explain the
influence of these processes because the most dominant process among the
three depends on redshift and photon energy, and the shape of the
spectrum of secondary particles is very sensitive to primary photon
energy.  At high energy, however, the fraction of heat, excitation and
ionization become constant values just like electron.  As high energy
photons are closely related with electrons, these constants will become the
almost same value in either case.

\begin{figure}[t]
 \begin{center}
  \includegraphics[width=0.6\linewidth]{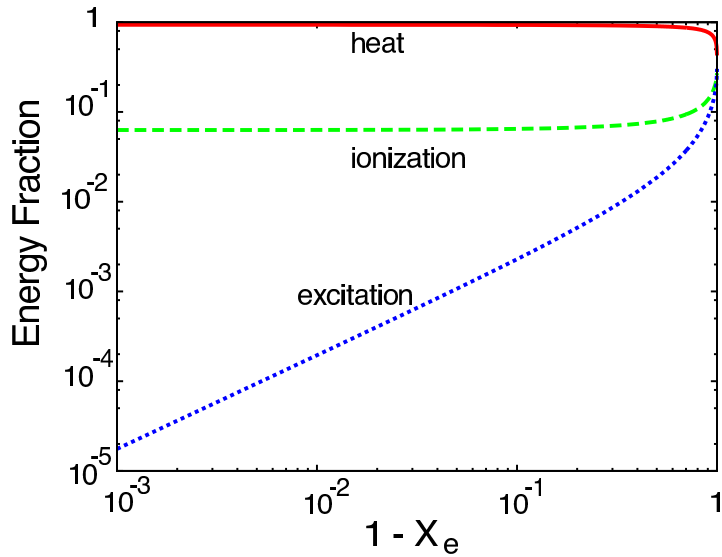}
  \caption{Electron and photon energy degradation with $1+z=1000$. 
  Solid line represents $\chi_h$, dashed line represents $\chi_{i}$ and
  dotted line represents $\chi_{ex}$.}
  \label{fig:fraction_xe}
 \end{center}
\end{figure}

\begin{figure}[t]
 \begin{center}
  \includegraphics[width=0.6\linewidth]{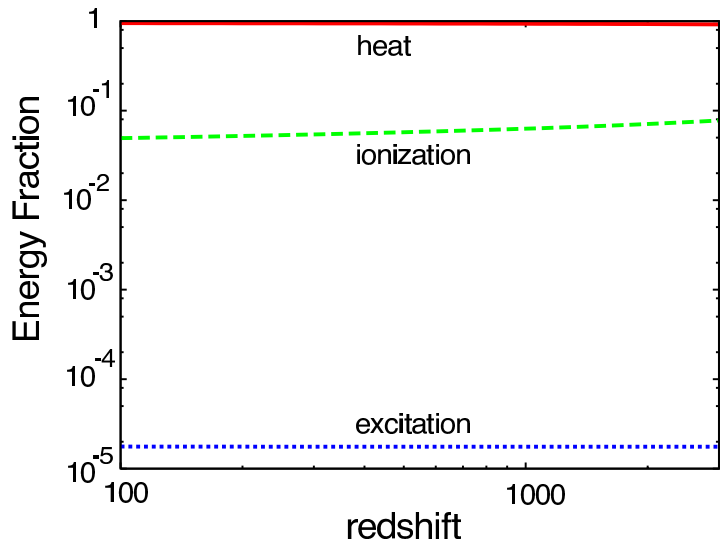}
  \caption{Electron and photon energy degradation with $x_e=10^{-3}$. 
  Solid line represents $\chi_h$, dashed line represents $\chi_{i}$ and
  dotted line represents $\chi_{ex}$.}
  \label{fig:fraction_z}
 \end{center}
\end{figure}

To see how ionization fraction $x_e$ and redshift $z$ have influence on
the energy degradation, we plot the fractions of heat, ionization and
excitation as function of $1-x_e$ and $1+z$ in
Figs.~\ref{fig:fraction_xe} and \ref{fig:fraction_z}, repectively.  Here
we have taken $10^{12}$~eV for the initial energy of electrons or
photons.  This is because all the fraction become constant values
at this energy.  Unless ionization fraction is very small, $\chi_h$ and
$\chi_i$ are almost independent of it.  This is because $\chi_i$ is
mainly determined by photoionization due to low energy secondary 
photons in this case and the collision
frequency of photoionization is much larger than that of Compton
scattering at low energy[Fig.~\ref{fig:frequency_photon}].
Therefore the effect of the change of
ionization fraction is almost irrelevant.  However, the fraction of
excitation $\chi_e$ is very sensitive to ionization fraction since the
collision frequency of electron impact excitation is in proportion to
$1-x_e$ while the competing processes (inverse Compton and Coulomb loss)
are independent of $1-x_e$.  The fraction of heat $\chi_h$ does not
depend on $1-x_e$ since $\chi_i$ is independent of $1-x_e$ and $\chi_{ex}$
is very small quantity. (Please notice that $\chi_i +\chi_{ex}+\chi_h =
1$.) When the ionization fraction is very small, the effect of Coulomb
scattering is weakened and these fractions become the same order.
Unlike ionization fraction, the change of redshift seems to have no
influence on the energy degradation.  This is because the change of
redshift effects which process is dominant at high energy, but is almost
irrelevant at low energy.  $\chi_i$ and $\chi_e$ are determined by 
secondary low energy particles and these fractions are nearly 
independent of redshift.

\section{Conclusions}
\label{sec:conclusion}

We have carried out detailed calculations of the fractions of the
initial energy of the injected electron or photon which are used to
heat, ionize and excite background plasma in the early universe.  In the
high energy limit ($E > 10^{12}$ GeV), we have shown that the fractions are
alomost independent of the initial energy.
Our calculations are valid up to TeV and can be applied to various
cosmological and astrophysical situations such as dark matter
decay/annihilation, which will be studied in a separete paper. 

\vskip 0.5cm

\noindent
{\it Acknowledgement}: This work was supported in part by the
Grant-in-Aid for Scientific Research from the Ministry of Education,
Science, Sports, and Culture of Japan,  No 14102004
(M.K.).  This work was supported by World Premier International
Research Center InitiativeiWPI Initiative), MEXT, Japan.
This work was also supported in part by JSPS-AF Japan-Finland
Bilateral Core Program.

\appendix

\section{Relevant Cross Sections}
\label{app:rel}

In this appendix, we show the cross sections and energy loss rates
adopted in this paper.

\subsection{Electron}

\subsubsection{Excitation of H\ }

We have adopted the almost same cross sections for atomic hydrogen and
Coulomb losses as~\cite{Dalgarno:1999}.  For the cross sections for the
electron impact excitation of atomic hydrogen, we have
adopted~\cite{Shimamura:1989}.  At high energy, we have used the Bethe
approximation reviewed
in~\cite{Inokuti:1971,Inokuti:1978,Kim:1971,Inokuti:1963}.

\begin{eqnarray}
 \sigma_{2p}({\rm NR}) & = & \frac{4\pi a^2_0}{T/{\rm Ry}}
  \left[M^2_{2p}\ln\left(\frac{4C_{2p}T}{{\rm Ry}}\right)
   +\frac{\gamma_{2p}}{T/{\rm Ry}}\right], 
  \label{eq:excitation_2p} \\
 \sigma_{2p}({\rm R}) & = & \frac{8\pi a^2_0}{mv^2/{\rm Ry}}M^2_{2p}
  \left[\ln\left(\frac{\beta^2}{1-\beta^2}\right)
   -\beta^2+\ln C_{2p}+11.2268\right],
\end{eqnarray}	
where $a_0$ is the Bohr radius, ${\rm Ry}$ is the Rydberg energy and $T
= mv^2/2$ represents the kinetic energy.  The numerical coefficients are
$\ln C_{2p} = -0.89704$, $\gamma_{2p} = 0.207985$ and $M^2_{2p} =
0.55493$.  $"{\rm NR}"$ and $"{\rm R}"$ mean non-relativistic and
relativistic respectively. 

\begin{eqnarray}
 \sigma_{2s}({\rm NR}) & = & \frac{4\pi a^2_0}{T/{\rm Ry}}
  \left[b_{2s}+\frac{\gamma_{2s}}{T/{\rm Ry}}\right], \\
 \sigma_{2s}({\rm R}) & = & \frac{8\pi a^2_0}{mv^2/{\rm Ry}}b_{2s},
  \label{eq:excitation_2s}
\end{eqnarray}	
where $b_{2s} = 0.11986$ and $\gamma_{2s} = -0.3125$.

\begin{eqnarray}
 \sigma_{n=3}({\rm NR}) & = & \frac{4\pi a^2_0}{T/{\rm Ry}}
  \left[M^2_{3}\ln\left(\frac{4C_{3}T}{{\rm Ry}}\right)
  \right], \\
 \sigma_{n=3}({\rm R}) & = & \frac{8\pi a^2_0}{mv^2/{\rm Ry}}M^2_{3}
  \left[\ln\left(\frac{\beta^2}{1-\beta^2}\right)
   -\beta^2+\ln C_{3}+11.2268\right],
  \label{eq:excitation_3}
\end{eqnarray}	
where $M^2_{3} = 8.8989\times10^{-2}$ and $\ln C_{3} = -0.2724$.

For the excitation to $n>3$, we subtract Eq.~(\ref{eq:excitation_2p})
$\sim$ Eq.~(\ref{eq:excitation_3}) from total excitation cross section.
\begin{eqnarray}
 \sigma_{ex}({\rm NR}) & = & \frac{4\pi a^2_0}{T/{\rm Ry}}
  \left[M^2_{ex}\ln\left(\frac{4C_{ex}T}{{\rm Ry}}\right)
   +\frac{\gamma_{ex}}{T/{\rm Ry}}\right], \\
 \sigma_{ex}({\rm R}) & = & \frac{8\pi a^2_0}{mv^2/{\rm Ry}}M^2_{ex}
  \left[\ln\left(\frac{\beta^2}{1-\beta^2}\right)
   -\beta^2+\ln C_{ex}+11.2268\right],
\end{eqnarray}	
where $\ln C_{ex} = -0.5780$,$\gamma_{ex} = -0.120575$ and $M^2_{ex} =
0.7166$.

\subsubsection{Ionization of H\ }

For the ionization cross section, we have adopted the following differential
cross section.
\begin{eqnarray}
 \frac{d\sigma_i(E,\epsilon)}{d\epsilon} = 
  \frac{A(E)}{1+(\epsilon/\bar{\epsilon})^2} 
  \hspace{1cm} {\rm for} \hspace{1cm} 0 \le \epsilon \le \frac{1}{2}(E-I),
  \label{diff_cs:ionization}	
\end{eqnarray}
where $E$ is the incident electron energy and $\epsilon$ is the energy
of the ejected electron.  We choose $\bar{\epsilon}=8$ eV.  In
Eq.(\ref{diff_cs:ionization}), a value of 2 is different from that of
2.1 originally suggested by~\cite{Opal:1971}.  This is why
Eq.(\ref{diff_cs:ionization}) can be analytically integrated.  Two
parameters $A(E)$ and $\bar{\epsilon}$ are related to the total
ionization cross section $\sigma_i(E)$ 
\begin{eqnarray}
 A(E) & = & \frac{\sigma_i(E)}{\bar{\epsilon}}[\tan^{-1}X(E)]^{-1}, \\
 X(E) & = & \frac{E-I}{2\bar{\epsilon}},
\end{eqnarray}	
where $I$ is the ionization potential.  The total ionization cross
section for atomic hydrogen had been measured by~\cite{Shah:1987} in the
range $14.6-4000$ eV.  Above 4000 eV, we used Bethe
approximation~\cite{Inokuti:1971,Kim:1971}.
\begin{eqnarray}
 \sigma_{i}({\rm NR}) & = & \frac{4\pi a^2_0}{T/{\rm Ry}}
  \left[M^2_{i}\ln\left(\frac{4C_{i}T}{{\rm Ry}}\right)
   +\frac{\gamma_{i}}{T/{\rm Ry}}\right], \\
 \sigma_{i}({\rm R}) & = & \frac{8\pi a^2_0}{mv^2/{\rm Ry}}M^2_{i}
  \left[\ln\left(\frac{\beta^2}{1-\beta^2}\right)
   -\beta^2+\ln C_{i}+11.2268\right],
\end{eqnarray}	
where $M^2_{i}=0.2834$, $\ln C_{i}=3.048$ and
$\gamma_{i}=-1.6294+\ln({\rm Ry}/T)$.

\subsubsection{$e^{-}$--H Collision\ }

For electron-hydrogen momentum transfer cross sections at
low energies, we have adopted the results
of~\cite{Bransden:1958,Dalgarno:1958}.  The momentum loss cross section
is described by
\begin{eqnarray}
 \sigma_{mt} = \frac{\pi a^2_0}{T/{\rm Ry}}
  \sum_{l=0}[3\sin^2(\eta^{-}_{l+1}-\eta^{-}_{l})
  +\sin^2(\eta^{+}_{l+1}-\eta^{+}_{l})]
\end{eqnarray}
where $\eta^{+}_l$ and $\eta^{-}_l$ are the phase shift computed
in~\cite{Rudge:1975,Das:1976}.  The cross section at 100, 200 and 300 eV
were calculated in~\cite{van Wyngaarden:1986}.  Cross sections at other
energies were derived by interpolation and extrapolation.  The energy
loss due to electron-hydrogen momentum transfer is described by
\begin{eqnarray}
 \left[\frac{-dE}{dt}\right]_{mt} = 
  \frac{2m_{e}E}{m_{p}}n_{H}v_{e}\sigma_{mt}(E),
\end{eqnarray}
where $m_{p}$ is the proton mass.

\subsubsection{Coulomb Collision\ }

Incident electrons lose their energies due to elastic collisions with
background electrons and photons.  Energy loss is dominated by electrons
at low energy since Coulomb cross sections are much larger than Compton
cross sections.  However, energy loss is dominated by photons at high
energy since the number density of photons is much larger than that of
electrons.  

For the energy loss due to Coulomb collisions with
background electrons, we have adopted the following analytical
formula~\cite{Swartz:1971}.
\begin{eqnarray}
 \left[\frac{-dE}{dt}\right]_{Cl} = 
  \frac{2.0\times 10^{-4}n^{0.97}_e}{E^{0.44}}
  \left(\frac{E-E_e}{E-0.53E_e}\right)^{2.36} {\rm eV}\cdot {\rm s}^{-1},
  \label{eq:Coulomb_loss}	
\end{eqnarray}
where $E$ is the incident electron energy in eV, $E_e$ is the background
electron energy in eV and $n_e$ is the electron number density in ${\rm
cm}^{-3}$.

\subsubsection{Inverse Compton Scattering\ }

An important quantity which characterizes the behavior of inverse
Compton scattering is $\gamma E_{CMB}$ (photon energy in the electron's
rest frame) where $E_{CMB}$ is the energy of CMB photon.  If $\gamma
E_{CMB}$ is much less than $m_e$, Thomson scattering approximation is
valid.  Otherwise, Klein-Nishina cross section should be used.

For inverse Compton scatterin with $\gamma
E_{CMB} \ll m_e$, the energy spectrum of scattered photon is obtained
by~\cite{Blumenthal:1970} in the limit $\beta \to 1$.  However, we
should keep $\beta$ so as not to spoil the validity of Thomson
approximation.  After some tedious calculations, the number of
collisions per unit time and unit scattered photon energy ($E_\gamma$)
is given by
\begin{eqnarray}
 \frac{d^2N}{dtdx} = \sigma_{T}cn(E_{CMB})dE_{CMB}f(x),
\end{eqnarray}
where $\sigma_T$ is the Thomson cross section and $n(E_{CMB})$ is the
differential number density of CMB photons and $x = E_\gamma/E_{CMB}$.
The expressions for $f(x)$ is given by
\begin{eqnarray}
 f(x) & = & \frac{3}{16\gamma^4\beta^4}
  \left[-(1+\beta^2)\left(\frac{1}{1+\beta}-\frac{x}{1-\beta}\right)
   +(x^2(1+\beta)-x(1-\beta))\right. \nonumber \\
 & - & \left.2x\ln\left(x\frac{1+\beta}{1-\beta}\right)\right]  
  \hspace{0.5cm} {\rm for} \hspace{0.5cm}
  \frac{1-\beta}{1+\beta} \le x \le 1, \\
 f(x) & = & \frac{3}{16\gamma^4\beta^4}
  \left[-(1+\beta^2)\left(\frac{x}{1+\beta}-\frac{1}{1-\beta}\right)
   +(x(1+\beta)-x^2(1-\beta))\right. \nonumber \\
 & - & \left.2x\ln\left(\frac{1}{x}\frac{1+\beta}{1-\beta}\right)\right]  
  \hspace{0.5cm} {\rm for} \hspace{0.5cm} 
  1 \le x \le \frac{1+\beta}{1-\beta}.
  \label{eq:cross_section_IC_Thomson}
\end{eqnarray}
The coefficient is determined so that $\int f(x)dx$ is equal to unity.
In the limit $\beta \to 1$, Eq.~(\ref{eq:cross_section_IC_Thomson})
corresponds to the result of~\cite{Blumenthal:1970}.  The number of
collisions per unit time and the energy loss rate can be easily
obtained.
\begin{eqnarray}
 \int \frac{d^2N}{dtdx}dx = \sigma_{T}cn(E_{CMB})dE_{CMB} 
  = \sigma_{T}cn_{CMB}, \\
 \int \frac{d^2N}{dtdx}(E_\gamma-E_{CMB})dx 
  =  \frac{4}{3}\sigma_{T}cn(E_{CMB})E_{CMB}dE_{CMB}\gamma^2\beta^2 
  \nonumber \\
 \hspace{4.4cm} = \frac{4}{3}\sigma_{T}cU_{CMB}\gamma^2\beta^2,
\end{eqnarray}	 
where $n_{CMB}$ and $U_{CMB}$ are the number and energy density of CMB
photons.

For inverse Compton scattering with $\gamma
E_{CMB} \ge m_e$, we should use Klein-Nishina cross section instead of
Thomson cross section.  The number of collisions per unit time and unit
scattered photon energy is given by~\cite{Jones:1968}
\begin{eqnarray}
 \frac{d^2N}{dtd\alpha'}  & = & 
  \frac{2\pi r^2_{e}c}{\alpha\gamma^2}
  \left[2q\ln q+(1+2q)(1-q)+\frac{1}{2}
   \frac{(4\alpha\gamma q)^2}{1+4\alpha\gamma q}(1-q)\right]
  \nonumber \\
 & \times & n(E_{CMB})dE_{CMB} 
  \hspace{0.5cm} {\rm for} \hspace{0.5cm} 
  \alpha \le \alpha' \le \frac{4\alpha\gamma^2}{1+4\alpha\gamma}
  \label{eq:cross_section_IC_KN}
\end{eqnarray}	
where $\alpha = E_{CMB}/m_e$, $\alpha' = E_\gamma/m_e$ and $q =
\alpha'/4\alpha\gamma^2(1-\alpha'/\gamma)$.
The number of collisions per unit time can be
obtained by integrating Eq.~(\ref{eq:cross_section_IC_KN}).  We shall
assume that Eq.~(\ref{eq:cross_section_IC_KN}) is valid for $0<q<1$,
even though Eq.~(\ref{eq:cross_section_IC_KN}) is quite invalid for
$0<q<1/4\gamma^2$.  The contribution from the region $0<q<1/4\gamma^2$
is $O(1/\gamma^2)$ and is negligible since $E_{CMB}$ is much less than
$m_e$~\cite{Jones:1968}.  The number of collisions per unit time is
given by~\cite{Zdziarski:1988}
\begin{eqnarray}
 \int \frac{d^2N}{dtd\alpha'}d\alpha' \simeq 
  \int^1_0 \frac{d^2N}{dtdq}dq = \sigma_T c \psi_1(s)n(E_{CMB})dE_{CMB},
\end{eqnarray}
where
\begin{eqnarray}
 \psi_1(s) = \frac{3}{2s^2}
  \left[\left(s+9+\frac{8}{s}\right)\ln(1+s)-8
   -\frac{2s+s^2}{2+2s}+4{\rm Li}_2(-s)\right],
\end{eqnarray}
and $s = 4\alpha\gamma$.  The function ${\rm Li}_2(x)$ is the
dilogarithm
\begin{eqnarray}
 {\rm Li}_2(x) \equiv -\int^{x}_0dz\frac{\ln(1-z)}{z}.
\end{eqnarray}
The energy loss rate is given by~\cite{Jones:1968}
\begin{eqnarray}
 \int E_\gamma \frac{d^2N}{dtd\alpha'}d\alpha' \simeq 
  \frac{4}{3}\sigma_T c\gamma^2\psi_2(s)n(E_{CMB})E_{CMB}dE_{CMB},
\end{eqnarray}
where
\begin{eqnarray}
 \psi_2(s)  & = & 
  \frac{9}{s^3}\left[\left(\frac{s}{2}+6+\frac{6}{s}\right)\ln(1+s)\right. 
  \nonumber \\
 & - & \left.\frac{6+13s+8s^2+11s^3/12}{(1+s)^2}+2{\rm Li}_2(-s)\right].
\end{eqnarray}

\subsection{Photon}

\subsubsection{Photoionization}

The absorption of X-rays and $\gamma$-rays is studied in detail
in~\cite{Zdziarski:1989}.  Incident photons are mainly absorbed by
hydrogen atoms and eject photoelectrons at low energies.  
The photoionization cross section for atomic hydrogen was reviewed
in~\cite{Heitler:1954,Lang:1999}.
\begin{eqnarray}
 \sigma_K({\rm NR}) & = & \frac{64\pi\sigma_{T}}{\alpha^3}
  \left(\frac{I}{h\nu}\right)^4
  \frac{\exp(-4\eta\cot^{-1}\eta)}{1-\exp(-2\pi\eta)}, \\
 \sigma_K({\rm R}) & = & \frac{3\sigma_{T}\alpha^4}{4}
  \left(\frac{m_e}{h\nu}\right)^5[\gamma^2-1]^{3/2} \nonumber \\
 & \times &
  \left[\frac{4}{3}+\frac{\gamma(\gamma-2)}{\gamma+1}
   \left(1-\frac{1}{2\gamma\sqrt{\gamma^2-1}}
    \ln\left(\frac{\gamma+\sqrt{\gamma^2-1}}{\gamma-\sqrt{\gamma^2-1}}\right)
   \right)\right]
\end{eqnarray}
where $h\nu$ is the incident photon energy, $\alpha$ is the
fine-structure constant, $I$ is ionization energy, $\eta =
1/\sqrt{h\nu/I-1}$ and $\gamma = (h\nu+m_e)/m_e$.  The cross sections
above are just halves of~\cite{Heitler:1954,Lang:1999}.  This is because
there is only one electron in K shell in the case of hydrogen.

\subsubsection{Compton Scattering}

Incident photons interact with background electrons through Compton
scattering.  If photon energy is sufficient small, the energy of recoil
electron is below the threshold energy of excitation and ionization of
atomic hydrogen.  Therefore the energy transferred to the recoil
electron can be regarded as heating.  Besides, a photon loses only a
small fraction of its energy per scattering.  
The energy loss due to
Compton scattering is described by~\cite{Zdziarski:1989}
\begin{eqnarray}
 \left[\frac{-dE}{dt}\right]_{Compton} = m_{e}n_{e}c\sigma_{T}x^2g(x),
\end{eqnarray}
where $x = h\nu/m_{e}$ and
\begin{eqnarray}	
 g(x) & = & \frac{3}{8}\left[\frac{(x-3)(x+1)}{x^4}\ln(1+2x)\right.
  \nonumber \\
 & +
  &\left.\frac{2(3+17x+31x^2+17x^3-\frac{10}{3}x^4)}{x^3(1+2x)^3}\right], 
  \\
 & \simeq & 1, \hspace{1cm} {\rm for} \hspace{0.5cm}	 x \ll 1; \\
 & \simeq & \frac{3}{8x^2}\left(\ln2x-\frac{5}{6}\right), 
  \hspace{1cm} {\rm for} \hspace{0.5cm} x \gg 1;
\end{eqnarray}	
If photon energy is as large as the electron mass, a photon loses a
sizable fraction of its energy per scattering.  In this case, it is
necessary to calculate the energy distribution of the recoil electrons.
The cross section is given by the following Klein-Nishina formula:
\begin{eqnarray}
 \frac{d\sigma}{d\epsilon}(h\nu) & = & 
  \frac{3\sigma_{T}}{8}\frac{m_e}{(h\nu)^2}
  \left[\frac{h\nu}{\epsilon}+\frac{\epsilon}{h\nu}
   +\left(\frac{m_e}{\epsilon}-\frac{m_e}{h\nu}\right)^2-2
   \left(\frac{m_e}{\epsilon}-\frac{m_e}{h\nu}\right)\right] 
  \nonumber \\
 & {\rm for} & \hspace{1cm} \frac{m_e}{m_e+2h\nu}h\nu \le \epsilon \le h\nu,
\end{eqnarray}	
where $\epsilon$ is the scattered photon energy.

\subsubsection{Pair Creation}

If photon energy is larger than $2m_e$, it is possible to create
electron-positron pair.  The energy and momentum conservation, however, are
only possible if another particle is present.  
The differential cross
section for pair creation in nuclei is given by the Bethe-Heitler
formula~\cite{Heitler:1954,Motz:1969} :
\begin{eqnarray}
 \frac{d\sigma}{dE_{+}} & = & 
  \alpha r^2_e\frac{p_{+}p_{-}}{E^2_\gamma}
  \left\{-\frac{4}{3}-2E_{+}E_{-}\frac{p^2_{+}+p^2_{-}}{p^2_{+}p^2_{-}}
   +m^2_e\left(\frac{E_{+}l_{-}}{p^3_{-}}+\frac{E_{-}l_{+}}{p^3_{+}}
	  -\frac{l_{+}l_{-}}{p_{+}p_{-}}\right)\right. \nonumber \\
 &+&
  L\left[\frac{E^2_\gamma}{p^3_{+}p^3_{-}}(E^2_{+}E^2_{-}+p^2_{+}p^2_{-})
    -\frac{8}{3}\frac{E_{+}E_{-}}{p_{+}p_{-}}
    -\frac{m^2_eE_\gamma}{2p_{+}p_{-}}\right. \nonumber \\
 &\times& \left.\left.\left(\frac{E_{+}E_{-}-p^2_{-}}{p^3_{-}}l_{-}
  +\frac{E_{+}E_{-}-p^2_{+}}{p^3_{+}}l_{+}
  +\frac{2E_\gamma E_{+}E_{-}}{p^2_{+}p^2_{-}}\right)\right\}\right],
\end{eqnarray}
where 
\begin{eqnarray}
 p_{\pm} = \sqrt{E^2_{\pm}-m^2_e}, \\
 L = \ln \frac{E_{+}E_{-}+p_{+}p_{-}+m^2_e}{E_{+}E_{-}-p_{+}p_{-}+m^2_e}, \\
 l_{\pm} = \ln \frac{E_{\pm}+p_{\pm}}{E_{\pm}-p_{\pm}},
\end{eqnarray}
and $E_{\pm}$ is the energy of positron (electron) energy.  The
analytical expression for cross section is given
by~\cite{Motz:1969,Maximon:1968}
\begin{eqnarray}
 \sigma & = & \alpha r^2_e
  \left\{\frac{28}{9}\ln 2k-\frac{218}{27}+\left(\frac{2}{k}\right)^2
   \left[6\ln 2k-\frac{7}{2}+\frac{2}{3}\ln^3 2k-\ln^2 2k\right.\right. 
  \nonumber \\
 &-& \left.\frac{\pi^2}{3}\ln 2k+\frac{\pi^2}{6}+2\zeta(3)\right]
  -\left(\frac{2}{k}\right)^4\left[\frac{3}{16}\ln 2k+\frac{1}{8}\right] 
  \nonumber \\
 &-& \left.\left(\frac{2}{k}\right)^6
  \left[\frac{29}{9\times256}\ln 2k-\frac{77}{27\times512}\right]
 +\cdots\right\} \hspace{0.5cm} {\rm for} \hspace{0.5cm} k > 4
\end{eqnarray}
where $k=E_\gamma/m_e$.  Convenient approximate formulas are given
by~\cite{Hough:1948} which are valid for $k\le20$,
\begin{eqnarray}
 \frac{d\sigma}{dx} = \alpha r^2_e\phi_0 z[1+0.135(\phi_0-0.52)z(1-z^2)] ,
  \label{eq:pair_production_Hough}
\end{eqnarray}
where
\begin{eqnarray}
 x & = & \frac{E_{+}-m_e}{E_\gamma-2m_e}, \\
 z & = & 2\sqrt{x(1-x)},
\end{eqnarray}
and $\phi_0$ is the differential cross section for equal partition of
energy, $E_{+}=E_{-}=E_\gamma/2$.  The second term in the square bracket
should be dropped when it becomes negative (below $k = 4.2$).  $\phi_0$
is given by
\begin{eqnarray}
 \phi_0 = (1-\gamma_1)\left[\frac{1}{3}(4-\gamma^2_1)(L_1-1)
   -\gamma^2_1\alpha_1(\alpha_1-1)-\gamma^4_1\alpha_1(L_1-\alpha_1)\right]
\end{eqnarray}
where
\begin{eqnarray}
 \gamma_1 & = & \frac{2}{k}, \\
 L_1 & = & \frac{2}{1-\gamma^2_1}\ln\left(\frac{k}{2}\right), \\
 \alpha_1 & = & \frac{1}{\sqrt{1-\gamma^2_1}}\ln\left[\frac{k}{2}
 +\sqrt{\left(\frac{k}{2}\right)^2-1}\right].
\end{eqnarray}
We get from Eq.~(\ref{eq:pair_production_Hough}) for the total cross section
\begin{eqnarray}
 \sigma & = & \frac{\pi}{4}\alpha r^2_e\phi_0 
  \hspace{3.3cm} {\rm for } \hspace{0.5cm} k < 4.2, \\
 & = & \alpha r^2_e(0.776\phi_0+0.018\phi^2_0)
  \hspace{0.5cm} {\rm for } \hspace{0.5cm} k > 4.2.
\end{eqnarray}

\subsubsection{Photon-Photon Scattering}

If the photon energy is below the effective threshold energy of the
double photon pair creation, photon-photon scattering
($\gamma\gamma\to\gamma\gamma$) process becomes significant.  
The photon-photon
scattering rate for $E_\gamma E_{CMB} \le m^2_e$ is given
by~\cite{Svensson:1990}
\begin{eqnarray}
 P(E_\gamma) = 3.33\times10^{11}\left(\frac{T_{CMB}}{m_e}\right)^6
  \left(\frac{E_\gamma}{m_e}\right)^3 {\rm s}^{-1}.
\end{eqnarray}
Normalized distribution of secondary photons of energy $E'_\gamma$,
$p(E'_\gamma,E_\gamma)$, is given by
\begin{eqnarray}
 p(E'_\gamma,E_\gamma) = \frac{20}{7}\frac{1}{E_\gamma}
 \left[1-\frac{E'_\gamma}{E_\gamma}
 +\left(\frac{E'_\gamma}{E_\gamma}\right)^2\right]^2
 \hspace{0.5cm} {\rm for} \hspace{0.5cm} 0 \le E'_\gamma \le E_\gamma. 
\end{eqnarray}
The distribution $p(E'_\gamma,E_\gamma)$ satisfies
\begin{eqnarray}
 \frac{1}{2}\int^{E_\gamma}_{0}p(E'_\gamma,E_\gamma)dE'_\gamma & = & 1, \\
 \int^{E_\gamma}_{0}p(E'_\gamma,E_\gamma)E'_\gamma dE'_\gamma & = & E_\gamma.
\end{eqnarray}	
The above formulas are not valid for a larger value of $E_\gamma$.
However, photon-photon scattering is not significant for high energy
photons since double photon pair creation is the dominant process.
Therefore, instead of using exact formulas, we simply neglect
photon-photon scattering for $E_\gamma E_{CMB}>m^2_e$.

\subsubsection{Double Photon Pair Creation\ }

For high energy photon, double photon pair creation ($\gamma\gamma\to
e^{+}e^{-}$) is the dominant process.  The total cross section for
double photon pair creation is given by~\cite{Gould:1967}
\begin{eqnarray}
 \sigma = \frac{1}{2}\pi r^2_e(1-\beta^2)
  \left[(3-\beta^4)\ln\frac{1+\beta}{1-\beta}-2\beta(2-\beta^2)\right],
\end{eqnarray}
where $\beta$ is the electron (positron) velocity in the center-of-mass
system.  The relationship between $\beta$ and $E_\gamma$, $E_{CMB}$ and
$\theta$ which is the angle between the momenta of the colliding photons
is easily obtained:
\begin{eqnarray}
 \beta & = & \sqrt{1-\frac{1}{s}}, \\
 s & = & \frac{E_\gamma E_{CMB}}{2m^2_e}(1-\cos\theta).
\end{eqnarray}	 
Clearly, the threshold energy for double photon pair production is
$E_\gamma=m^2_e/E_{CMB}$, head-on photon collision ($\theta=\pi,s=1$).
For calculation of the absorption probability, we should average the
above cross section over the distributions for isotropically distributed
photons~\cite{Gould:1967,Zdziarski:1988},
\begin{eqnarray} 
 \sigma_{ave} & = & \frac{1}{2}
  \int^{1-2m^2_e/E_\gamma E_{CMB}}_{-1}(1-\cos\theta)\sigma d\cos\theta 
  \nonumber \\
 & = & \frac{3}{8}\sigma_T\left(\frac{m^2_e}{E_\gamma E_{CMB}}\right)^2
  \left[\frac{1+2v+2v^2}{1+v}\ln\omega-2\sqrt{\frac{v}{1+v}}(1+2v)\right. 
  \nonumber \\
 & + & \left. 2\ln^2(1+\omega)-\ln^2\omega+4{\rm Li}_2
 \left(\frac{1}{1+\omega}\right)-\frac{\pi^2}{3}\right],
\end{eqnarray}
where
\begin{eqnarray}
 v & = & \frac{E_\gamma E_{CMB}}{m^2_e}-1 > 0, \\
 \omega & = & \frac{\sqrt{1+v}+\sqrt{v}}{\sqrt{1+v}-\sqrt{v}}.
\end{eqnarray}
Differential spectra of electrons and positrons are given
by~\cite{Agaronyan:1983}
\begin{eqnarray}
 \frac{d\sigma}{dE_e} & = & 
  \frac{\pi r^2_em^4_e}{4E^3_\gamma E^2_{CMB}}
  \left[\frac{4E^2}{E_eE_p}\ln \frac{4E_{CMB}E_eE_p}{m^2_eE}
   -\frac{8E_{CMB}E}{m^2_e}\right. \nonumber \\
  & + & \left.\frac{2(2E_{CMB}E-m^2_e)E^2}{m^2_eE_eE_p}
  -\left(1-\frac{m^2_e}{E_{CMB}E}\right)\frac{E^4}{E^2_eE^2_p}\right],
\end{eqnarray}
where $E_e$($E_p$) is the energy of electron (positron) and $E$ is the
total energy, $E=E_e+E_p=E_\gamma+E_{CMB}$.  The limits of the variation
of $E_e$($E_p$) is given by
\begin{eqnarray}
 \frac{E}{2}\left(1-\sqrt{1-\frac{m^2_e}{E_{CMB}E}}\right) 
  \le E_e \le \frac{E}{2}\left(1+\sqrt{1-\frac{m^2_e}{E_{CMB}E}}\right).
\end{eqnarray}

\subsection{Positronium}

Here, we show the energy spectrum from three-photon annihilation of
positronium.  The energy spectrum is continuous, as allowed by
conservation of momentum.  It has been calculated in~\cite{Ore:1949}
with photon energy $\eta$ normalized by electron mass,
\begin{eqnarray}
 F(\eta) & =  & \frac{2}{\pi^2-9}
  \left[\frac{\eta(1-\eta)}{(2-\eta)^2}-\frac{2(1-\eta)^2}{(2-\eta)^3}
  \ln(1-\eta)+\frac{2-\eta}{\eta}\right. \nonumber \\
 & + & \left.\frac{2(1-\eta)}{\eta^2}\ln(1-\eta)\right]. 
  \hspace{0.5cm} {\rm for} \hspace{0.5cm} 0 \le \eta \le 1
\end{eqnarray}
The function $F(\eta)$ is normalized so that
\begin{eqnarray}
 \int^1_{0}d\eta F(\eta) = 1.
\end{eqnarray}

\end{document}